\documentclass[11pt]{article}

\usepackage{amsmath, amssymb}
\usepackage{comment}
\usepackage{graphicx}
\usepackage{bm}
\usepackage[version=4]{mhchem}
\usepackage{float}
\usepackage{xcolor}
\usepackage{hyperref}
\usepackage{gensymb}
\usepackage[a4paper, margin=2.5cm]{geometry}
\usepackage{authblk}  
\usepackage[numbers,sort&compress]{natbib}
\usepackage{caption}
\usepackage{setspace}

\DeclareCaptionType{extfigure}[Supplementary Fig.][List of Supplementary Figures]

\usepackage[labelfont=bf]{caption}

\hypersetup{
  colorlinks   = true,
  urlcolor     = blue,
  linkcolor    = blue,  
  citecolor    = blue   
}

\title{\bfseries Unveiling the Interplay of Charge and Magnetic Excitations in HgBa$_2$Ca$_2$Cu$_3$O$_{8+\delta}$}

\author[1,2]{Karn Rongrueangkul}
\author[3]{Martina Fedele}
\author[3]{Leonardo Martinelli\thanks{Present address: Physik-Institut, Universität Zürich, Winterthurerstrasse 190, CH-8057 Zürich, Switzerland}}
\author[3]{Giacomo Merzoni}
\author[4]{Roberto Sant}
\author[4]{Nicholas B. Brookes}
\author[5]{Dorothée Colson}
\author[6]{Alain Sacuto}
\author[7]{Götz Seibold}
\author[8]{Sergio Caprara}
\author[3]{Marco Moretti Sala}
\author[3,9]{Giacomo Ghiringhelli\thanks{e-mail: \protect\href{mailto:giacomo.ghiringhelli@polimi.it}{giacomo.ghiringhelli@polimi.it}}}
\author[1,10]{Riccardo Arpaia\thanks{e-mail: \protect\href{mailto:riccardo.arpaia@unive.it}{riccardo.arpaia@unive.it}}}

\affil[1]{Quantum Device Physics Laboratory, Department of Microtechnology and Nanoscience, Chalmers University of Technology, SE-41296 Göteborg, Sweden}
\affil[2]{Wallenberg Initiative Materials Science for Sustainability, Department of Microtechnology and Nanoscience, Chalmers University of Technology, SE-41296 Göteborg, Sweden}
\affil[3]{Dipartimento di Fisica, Politecnico di Milano, piazza Leonardo da Vinci 32, I-20133 Milano, Italy}
\affil[4]{ESRF, The European Synchrotron, 71 Avenue des Martyrs, CS 40220, F-38043 Grenoble, France}
\affil[5]{Service de Physique de l'Etat Condensé, CEA Saclay, IRAMIS, SPEC (CNRS URA 2464), F-91191, Gif sur Yvette, France}
\affil[6]{Université Paris Cité, Matériaux et Phénomènes Quantiques, UMR CNRS 7162, Bâtiment Condorcet, F-75205, Paris Cedex 13, France}
\affil[7]{Institut f\"ur Physik, BTU Cottbus-Senftenberg, D-03013 Cottbus, Germany}
\affil[8]{Dipartimento di Fisica, Università di Roma ``La Sapienza'', P.le Aldo Moro 5, I-00185 Roma, Italy}
\affil[9]{CNR-SPIN, Dipartimento di Fisica, Politecnico di Milano, I-20133 Milano, Italy}
\affil[10]{Department of Molecular Sciences and Nanosystems, Ca’ Foscari University of Venice, I-30172 Venice, Italy}

\date{}
\begin{document}
\newcommand{\gm}[1]{\textcolor{magenta}{#1}}
\maketitle
\newpage
\begin{abstract}
   \textbf{Unraveling the mechanism that binds electrons into Cooper pairs in cuprate high-temperature superconductors remains one of the most fundamental challenges in condensed-matter physics. While both magnetic interactions and lattice vibrations are known to govern key electronic properties, their possible cooperation has never been directly observed. We investigate HgBa$_2$Ca$_2$Cu$_3$O$_{8+\delta}$ (Hg1223)—the cuprate with the highest $T{\mathrm{c}}$ at ambient pressure—as a magnifying glass to probe the possible entwining of the excitations at the core of the pairing. 
   Using resonant inelastic X-ray scattering, we find that the charge response is dominated by dynamic charge density fluctuations (CDF) extending up to several hundred meV, where magnetic excitations reside. At the same momentum where CDF are most intense, the paramagnon energy exhibits a pronounced softening, revealing a strong interplay among charge, lattice, and spin excitations. Our results point to a cooperative mechanism in which dynamic charge fluctuations mediate the coupling between lattice, charge and spin degrees of freedom—shedding new light on the fundamental origin of high-$T{\mathrm{c}}$ superconductivity.}
\end{abstract}

\newpage

\section{\label{sec:introduction}Introduction}

In strongly correlated oxides, a variety of emergent phenomena arises from the interplay between different excitations and competing instabilities, which determine the ground state of these materials~\cite{dagotto2005complexity, keimer2015quantum, rau2016spin}. Among them, the coupling between spin and lattice degrees of freedom plays a central role. In multiferroics, for instance, spin–lattice interactions underpin a broad range of effects, from the thermal Hall response~\cite{ideue2017giant} to multiferroicity itself~\cite{fiebig2016evolution} and temperature-driven ferroelectric transitions~\cite{cazorla2017multiple}.  

In cuprate high-temperature superconductors (HTS), a coupling between charge, lattice and spin degrees of freedom has long been invoked to explain the still elusive pairing mechanism \cite{fradkin2015colloquium}. This is natural, since the superconducting dome is framed by a Mott insulating state—highlighting the role likely played by magnetic interactions in the pairing—and an overdoped Fermi-liquid regime where superconductivity appears more conventional, framed within a dirty-$d$-wave extension of the Bardeen–Cooper–Schrieffer (BCS) theory which entails a crucial contribution from electron–phonon coupling \cite{keimer2015quantum}. Several works have indeed reported a correlation between the antiferromagnetic exchange interaction $J$ and $T_{\mathrm{c}}$ \cite{munoz2000accurate, wang2022paramagnons}, although no general consensus has been reached, as this relation is not universal among different cuprate families and $J$ is nearly doping independent. The lattice, on the other hand, is clearly affected by doping: oxygen-related phonon modes progressively soften with increasing hole concentration up to slightly above the optimal level, where this effect reaches its maximum 
\cite{reznik2006electron, fedele2026electron}. The emerging picture suggests that magnetism is essential, yet, most likely, does not act alone.

Despite numerous indications that the properties of high-$T_{\mathrm{c}}$ cuprates depend on both lattice interactions and magnetic fluctuations, a direct observation of a genuine spin–phonon coupling (SPC) 
has so far remained elusive. Theoretically, such coupling could account for several key phenomena, including superconductivity itself, isotope effects, and the temperature dependence of the pseudogap \cite{jarlborg2009spin}. Recently, thermal Hall conductivity measurements have revealed 
chiral phonons — i.e., phonons carrying finite angular momentum — in the pseudogap regime of cuprates \cite{grissonnanche2020chiral}. 
The presence of such excitations, known to be associated with strong SPC in multiferroics, reinforces the need to further explore spin-phonon coupling in cuprates.
To elucidate the interplay among electronic, lattice, and magnetic excitations, we investigated HgBa$_2$Ca$_2$Cu$_3$O$_{8+\delta}$ (Hg1223). As a record-high-$T_c$ compound, this system provides an ideal platform for probing the fundamental coupling mechanisms, under the assumption that in this case even the most elusive interactions can exceed the experimental sensitivity. 
Hg-based cuprates provide an ideal platform for investigating intrinsic pairing interactions, as the first three members of the Hg homologous series, HgBa$_2$CuO$_{4+\delta}$, HgBa$_2$CaCu$_2$O$_{6+\delta}$, and HgBa$_2$Ca$_2$Cu$_3$O$_{8+\delta}$, exhibit the highest ambient-pressure $T_c$ values among single-, double-, and triple-layer cuprates, respectively.

Hg1223, in particular, exhibits the highest ambient-pressure $T_{\mathrm{c}}$ among all cuprates. Its trilayer structure hosts spin and charge excitations~\cite{oliviero2022magnetotransport,loret2019intimate}, together with the largest superconducting gap ever observed in a cuprate~\cite{wen2025unprecedentedly}. 

We employed resonant inelastic X-ray scattering (RIXS) at the Cu $L_3$ edge to simultaneously probe charge, lattice, and magnetic excitations, a measurement that had not been reported on Hg1223 before. By mapping the momentum dependence of the charge order with unprecedented energy resolution, we find that the signal mainly arises from dynamic charge-density fluctuations (CDF), which we recently showed to be closely linked to electron-phonon coupling and superconductivity~\cite{fedele2026electron}. Remarkably, we detect a pronounced softening of the paramagnon energy at the same momentum where CDF are most intense and show a high-energy tail up to several hundreds of meV. 
CDF thus act as a bridge between phonons and spin excitations. The observation of a pronounced spin–lattice entanglement in this system sheds new light on the cooperative interactions underlying high-$T_{\mathrm{c}}$ superconductivity.

\section{\label{sec: exp res}Results}
\subsection*{Choice of doping and measurement conditions}\label{material}
The choice of the doping level is crucial in the case of Hg1223, since the unit cell contains three CuO$_2$ layers with distinct local environments. The inner CuO$_2$ plane (IP) lacks apical oxygens, while the two outer planes (OPs) exhibit pyramidal CuO$_5$ coordination and are directly influenced by the adjacent HgO$_\delta$ charge reservoir layers (see Fig.~\ref{fig:figure1}(a)). Nuclear magnetic resonance \cite{kotegawa2001nmr, julien1996spin} and angle-resolved photoemission \cite{ideta2010enhanced, horio2025enhanced} studies have shown that this structural asymmetry gives rise to a charge imbalance, whereby the IP remains weakly doped whereas the OPs are more strongly doped. As a consequence, the nominal doping of a crystal reflects an average between two highly doped OPs and a poorly doped IP -- an imbalance which would complicate the interpretation of the excitations observed by RIXS. For this reason, we have chosen to work with a sample at a doping level of $p \approx 0.12$. At this underdoped composition, Cu Knight shift measurements demonstrated that the 
the OP–IP doping imbalance is minimized, while it increases progressively with higher carrier concentration \cite{kotegawa2001nmr}.

We have therefore investigated a high-quality single crystal with the desired doping level ($T_{\mathrm{c}} = 112$ K), grown by a self-flux method (see Methods for additional details) \cite{loret2017crystal}. 
RIXS measurements at the Cu $L_{3}$ edge ($\approx 932$ eV) were performed at the ID32 beamline of the European Synchrotron Radiation Facility (ESRF) \cite{Brookes}. The incident x-rays were $\sigma$-polarized, which provides a reasonable sensitivity to magnetic excitations while enhancing the cross sections of charge and lattice excitations. RIXS spectra were acquired by fixing the scattering angle at $2\theta = 149.5^\circ$ and spanning a broad momentum range (0.04–0.48 r.l.u.) along the $(H,0)$ and $(H,H)$ directions. Ultrahigh-resolution spectra (32 meV, stars in Fig.~\ref{fig:figure1}b) were collected at 110 and 300 K to resolve the lineshape of charge, lattice, and magnetic excitations, while medium-resolution data (59 meV, yellow bar in Fig.~\ref{fig:figure1}b) were acquired between 20~K and room temperature to track the temperature dependence of the quasielastic intensity and paramagnons.

\subsection*{Charge modulations and lattice excitations in Hg1223}\label{CO}

Here, by charge modulations we refer to the two components of charge order: the quasi-static charge-density waves (CDW) \cite{ghiringhelli2012long, chang2012direct}, known to compete with superconductivity, and the dynamic charge-density fluctuations (CDF), with meV-scale energies, which appear to be more deeply entwined with the superconducting state \cite{arpaia2021charge}. By lattice excitations we refer instead to the bond-stretching (BS) phonons, which at the Cu $L_3$ edge can be resolved more clearly than other low-energy modes. 

To disentangle these contributions, we analyzed the ultrahigh-resolution spectra (Fig.~\ref{fig:figure1}(c)). Each spectrum along the $(H,0)$ and $(H,H)$ directions was fitted below 1 eV with four narrow Gaussian peaks (elastic/CDW, CDF, BS phonons, and phonon overtones at increasing energy loss), a damped harmonic oscillator (DHO) for the magnetic excitations, and a linear background for the particle-hole continuum (Fig.~\ref{fig:figure1}(d), see Methods  for additional details).

Focusing on the low-energy region along $(H,0)$ at 110~K, we find that the intensity of the elastic peak grows when approaching the $\Gamma$ point, with no distinct features at finite momentum (Fig.~\ref{fig:figure2}(a)). By contrast, a CDF peak emerges at $q_{\mathrm{CDF}} = 0.30$~r.l.u., with a characteristic energy of about 15 meV (Fig.~\ref{fig:figure2}(b)). Finally, the BS phonon displays a pronounced anomaly: its energy softens significantly at $q_{\mathrm{CDF}}$ (Fig.~\ref{fig:figure2}(c)), and its intensity exceeds the expected dependence proportional to  $\sin^2(\pi q)$ (Fig.~\ref{fig:figure2}(d)) \cite{BraicovichPhon}.
Along $(H,H)$, on the contrary,  both the CDF signal and the phonon anomaly disappear (see Supplementary Fig. \ref{fig:HH}).
At 300~K  the results are very similar to those at 110~K in both directions (see Supplementary Fig.~\ref{fig:fit300}). 

Two key aspects must be underlined.

(i) Although at $p \approx 1/8$ the CDW intensity reaches a maximum in most cuprates, \textit{in Hg1223, the RIXS data show no clear evidence of a static CDW, with the charge order response mainly arising from finite-energy CDF}. This is consistent with the phase diagram of Hg1223, where the superconducting dome, as reconstructed from the available literature \cite{loret2017crystal,cohn1999, Passos2006}, is perfectly parabolic (Fig.~\ref{fig:figure1}(b)).
The absence of $T_{\mathrm{c}}$ anomalies in the underdoped region indicates negligible competition with CDW, in turn suggesting a very weak CDW correlations. 

(ii) Despite the absence of a clear quasi-static CDW peak in our RIXS spectra, the BS phonon softening is not only present, but even more pronounced than in other cuprates at comparable doping and temperature \cite{lee2021spectroscopic, wang2021charge,  fedele2026electron}. This further confirms, on the one hand, that CDF are primarily responsible for the observed phonon renormalization. On the other hand, \textit{since the electron–phonon coupling is directly proportional to this softening} \cite{wang2021charge, fedele2026electron}, \textit{our results indicate that it must be particularly strong in Hg1223}.

\subsection*{Magnetic excitations in Hg1223}\label{PM}

In Fig.~\ref{fig:figure2}(e) we show the high-resolution spectra at $T=110$~K after removing the non-magnetic contributions determined from the aforementioned fits. This procedure allows us to visualize the full momentum dispersion of the paramagnons. The energies corresponding to the maximum of the DHO peak, $\omega_{\mathrm{max}}$, are plotted in Fig.~\ref{fig:figure2}(f) along both the $(H,0)$ and $(H,H)$ directions. We first fitted this dispersion using a simple nearest-neighbour Heisenberg model with one effective in-plane exchange parameter $J_{\parallel}$ and  interlayer coupling $J_{\perp}$ (see dashed line in Fig.~\ref{fig:figure2}(f), and Methods for additional details). The resulting $J_{\parallel} \approx 180$~meV, mainly constrained by the magnetic Brillouin zone boundary (MBZB) paramagnon energy and known to be doping independent~\cite{LeTacon, dean2013persistence, jia2014persistent, PengPRB}, is, to our knowledge, the highest value reported for cuprates. In Fig.~\ref{fig:figure2}(h) we compare this value with those obtained, 
using the same analysis, for Hg1201 and Hg1212 from the data of Wang \emph{et al.}~\cite{wang2022paramagnons}, plotting them as a function of the maximum critical temperature achievable in each compound. Remarkably, in agreement with previous reports~\cite{ofer2006magnetic, wang2022paramagnons}, the in-plane Heisenberg exchange energy scales linearly with $T_{\mathrm{c}}^{\mathrm{max}}$.

Away from the MBZB, however, the agreement between the experimental paramagnon dispersion and the simple Heisenberg fit is poor. In particular, the large experimental dispersion along the MBZB, $\Delta E_{\mathrm{MBZB}}$, calls for a refined description including long-range in-plane exchange interactions. Following Peng \emph{et al.}~\cite{PengInfluence}, we extended the Heisenberg model to include effective exchange terms up to fourth-nearest neighbours (see solid line in Fig.~\ref{fig:figure2}(f), and Methods for additional details). The resulting parameters are plotted in Fig.~\ref{fig:figure2}(i), together with those of the single- and double-layer Hg cuprates. The same linear relation with $T_{\mathrm{c}}$ that holds for $J_{\parallel}$ now applies to the antiferromagnetic $J_{1}^{\mathrm{eff}}$ term. This extended Heisenberg fit reproduces the dispersion along both $(H,0)$ and $(H,H)$ directions, with the exception of the momentum region around $(0.3,0)$, where the measured energy falls below the theoretical expectation. The same conclusion is reached when fitting the dispersion of the undamped energy $\omega_{0}$ of the paramagnons (Fig.~\ref{fig:figure2}(g)). 

Two key aspects must be underlined. 

(i) Our results on the trilayer confirm what has been observed in the single- and double-layer compounds of the Hg family, further reinforcing the robust connection between paramagnon energy and superconducting $T_{\mathrm{c}}$. At the same time, this 
observation is necessary but not sufficient: other cuprates, such as those of the 214 family, exhibit relatively low $T_{\mathrm{c}}$ despite high $J_{\parallel}$ values. \textit{This highlights both the importance of magnetic excitations for stabilizing the superconducting ground state and the fact that they alone cannot account for the pairing interaction}.  

(ii) \textit{Magnetic excitations show a softening at $q_{\mathrm{CDF}}$}. To probe the connection between paramagnons and CDF we have investigated their temperature dependence .


\subsection*{Common temperature dependence of CDF and paramagnon softening}\label{Tdep}

To probe the temperature dependence of charge order, we integrated the quasi-elastic intensity of each medium-resolution spectrum up to 35~meV. The resulting peaks as a function of temperature (Fig.~\ref{fig:figure3}(a)), centered at $q \approx 0.30$~r.l.u., confirm that charge order in Hg1223 is dominated by CDF. Both the peak height and the correlation length -- given by the inverse of the FWHM -- exhibit only a mild evolution with temperature: they remain essentially constant up to $T_{\mathrm{c}}$, then decrease slowly above the superconducting transition, and are still finite at 300~K (Fig.~\ref{fig:figure3}(b)–(c)). 

To investigate the temperature dependence of the paramagnons, we fitted the mid-infrared region of the medium-resolution spectra with the DHO profile. Figures~\ref{fig:figure3}(d)–(g) show the extracted energies $\omega_{\mathrm{max}}$ (circles) at different temperatures, together with fits based on linear spin-wave theory including exchange interactions up to fourth-nearest neighbours (dashed lines). Strikingly, the paramagnon softening -- defined as the deviation between the experimental $\omega_{\mathrm{max}}$ around $q_{\mathrm{CDF}}$ and the theoretical dispersion -- systematically decreases with increasing temperature. 
To visualize this effect, we plot this deviation as a 
positive contribution in Fig.~\ref{fig:figure3}(h). This representation 
highlights the close similarity between these peaks and those shown in 
Fig.~\ref{fig:figure3}(a), both in their momentum width and in their 
temperature evolution.  In particular, as shown in 
Fig.~\ref{fig:figure3}(i), the peak height grows upon cooling and saturates below $T_{\mathrm{c}}$. This temperature dependence mirrors that of CDF, indicating a common entwining with superconductivity and suggesting a strong coupling between charge and spin excitations. 

The main difference is that, while CDF persist up to room temperature, 
the paramagnon softening is almost fully suppressed at 300~K. This 
apparent discrepancy will be reconciled in the next section by resolving 
the full energy profile of the CDF signal.




\subsection*{Broad-in-energy CDF overlapping with paramagnons}\label{CDFbroad}

To further gain insight into the possible interaction between CDF and magnetic excitations, we investigated the actual energy profile of the CDF signal and its temperature dependence. In Fig.~\ref{fig:figure2} the CDF were fitted to a symmetric Gaussian; here we relax this constraint. We divided the ultra-high-resolution RIXS spectra at each momentum into adjacent energy intervals of 60~meV (Fig.~\ref{fig:figure4}(a)) and determined for each interval the integrated intensity. The momentum dependence of the integrated signal for both 110 and 300~K is shown in Fig.~\ref{fig:figure4}(b)–(e). At low temperature, besides the expected quasi-elastic peak in the $[-30,30]$~meV range—similar in shape and FWHM to the CDF peak extracted from the medium-resolution spectra of Fig.~\ref{fig:figure3}(a)—we observe additional intensity still centered at $q_{\mathrm{CDF}}$, whose amplitude decreases with increasing energy but remains detectable up to 210~meV. At high temperature, by contrast, a broadened quasi-elastic peak persists (Fig.~\ref{fig:figure4}(b)), consistent with the temperature dependence of Fig.~\ref{fig:figure3}(a)–(c), but the high-energy component above 90~meV vanishes. This indicates the presence of a high-energy tail of the CDF, with a steeper temperature dependence than the low-energy component. 

To resolve more finely the energy profile of the CDF peak, we repeated the same procedure using 60~meV-wide intervals, shifted by only 10~meV from each other. This yields a denser mapping of the momentum-dependent peaks (Supplementary Figs.~\ref{fig:Greven110}–\ref{fig:Greven300}). By fitting these peaks with gaussians and plotting their integrated areas as a function of energy, we obtained the spectrum shown in Fig.~\ref{fig:figure4}(f) for 110~K. The resulting peak is centered around 15~meV and broader than the instrumental resolution, consistent with the CDF features identified in Fig.~\ref{fig:figure2}. Remarkably, the peak is asymmetric, exhibiting a pronounced high-energy tail that extends up to $\sim$300~meV. The inset of Fig.~\ref{fig:figure4}(f) shows that all Gaussian peaks are centered at $q = q_{\mathrm{CDF}} = 0.30 \pm 0.01$~r.l.u. Only in the 60–120~meV range, where the integrated intensity is dominated by BS phonons, the same $q$ could be recovered only after subtracting their contribution (see Methods). \textit{Thus, at temperatures around $T_{\mathrm{c}}$, the CDF extend up to the energy range of paramagnons, favouring a strong interaction between charge and magnetic excitations}. 



Unlike the low-energy component, this high-energy tail becomes negligible at 300~K: the CDF peak acquires a symmetric energy profile, and no significant spectral weight remains above 100~meV (Supplementary Figs.~\ref{fig:Greven300}--\ref{fig:CDF300}). Notably, this is precisely the temperature at which the paramagnon softening also vanishes, thereby resolving the apparent inconsistency that emerges at 300~K when only the low-energy CDF contribution is considered.

This observation further supports the conclusion that CDF lie at the 
origin of the paramagnon softening, with their high-energy tail likely 
representing a key ingredient in the strong coupling between charge and 
spin degrees of freedom.


\section{\label{sec:discussion}Discussion}

A softening of the paramagnon dispersion at the charge-ordering wave vector has been previously reported in cuprates, most notably in La$_{2-x}$Ba$_x$CuO$_4$ (LBCO)~\cite{miao2017high}. In this material, such a phenomenon is not entirely unexpected: as a member of the 214 family, LBCO hosts static stripe order, where spin and charge excitations are inherently coupled. The observed softening has been attributed to CDW, whose strong intensity causes the nearly complete suppression of superconductivity at $p = 1/8$. In this framework, the spin--charge coupling, when mediated by charge excitations that compete with superconductivity, may become detrimental to pairing, thereby accounting for the relatively low $T_{\mathrm c}$ of LBCO despite its sizable antiferromagnetic exchange $J_{\parallel}$.

The situation in Hg1223 is markedly different. Here, the contribution from static CDW is almost negligible, charge excitations are dominated by CDF, and superconductivity reaches the highest critical temperature of any cuprate at ambient pressure. This raises a fundamental question: can a spin--charge coupling mediated by dynamical charge fluctuations, rather than by static charge order, promote rather than suppress superconductivity?

A paramagnon softening, though weaker, has been reported in Hg1201 and Hg1212~\cite{yu2020unusual, wang2022paramagnons}, together with a dynamic component of charge order extending to comparable energies~\cite{yu2020unusual}. 
To determine whether this anomaly of the magnetic response is connected to the energy profile of the CDF,  we have performed analyses, analogous to those presented in Figure~\ref{fig:figure4},  on YBa$_2$Cu$_3$O$_{7-\delta}$ (YBCO), a system that—despite extensive investigation—has never shown any paramagnon softening~\cite{LeTacon, le2013dispersive, PengInfluence}. At both $p = 0.12$, comparable to the doping of our Hg1223 sample, and $p = 0.19$, where the CDF intensity is maximal, the CDF peak in YBCO is symmetric and confined below 80~meV (see Methods and Supplementary Figs.~\ref{fig:YBCOUD}–\ref{fig:CDFYBCO}). The comparison with YBCO therefore indicates that the paramagnon softening observed in Hg1223 
does not arise from the mere presence of CDF, but specifically from their extension to high energies.

Why do CDF in Hg-based cuprates extend to such high energies? As we have recently shown~\cite{fedele2026electron}, CDF are intimately connected to the electron--phonon coupling (EPC): the larger the CDF spectral weight, the stronger the EPC and the more pronounced the phonon anomaly at $q_{\mathrm{CDF}}$, with all these effects peaking at $p = 0.19$, where superconductivity is strongest. In Hg1223, the EPC is expected to be particularly strong owing to the crystal-field environment and reduced screening that enhance coupling to out-of-plane vibrational modes, 
such as the $B_{1g}$ phonons of the planar oxygen atoms in the CuO$_2$ planes and the $A_{1g}$ phonons of the apical oxygen atoms~\cite{johnston2010systematic}.
Consistently,  recent RIXS measurements at the O~$K$ edge revealed apical phonons and their overtones extending to several hundred meV~\cite{hong2025dominant}, pointing to an exceptionally strong EPC involving the apical oxygen and producing a Franck--Condon--like envelope at high energies~\cite{ament2011determining}. In the charge channel, this results in a corresponding effect: CDF acquire a pronounced high-energy tail, as observed in our experiment.


How can CDF account for the paramagnon softening in view of the strong electron--phonon coupling observed in Hg1223? 
Within a simple phenomenological picture, which we discuss here for clarity (see Methods and Supplementary Fig.~\ref{fig1M} for additional details, together with a microscopic model of the paramagnon spectrum in the presence of strongly interacting CDF), the partial softening at $q_{\mathrm{CDF}}$ can be viewed as a precursor effect, arising from the slow dynamics of the CDF in the presence of strong electron--phonon coupling.  These fluctuations locally drive the system towards a nearly reconstructed lattice configuration, characterized by an emergent periodicity $a' = 2\pi/|q_{\mathrm{CDF}}|$, larger than the lattice spacing $a$, yet without establishing a true static reconstruction. We denote by $\omega_k$ and $\omega'_k$ the paramagnon dispersion (see Eq. \ref{eq:paradispersions} for details) in the original and reconstructed states, respectively, and, for simplicity, consider a one-dimensional case. We impose that $\omega_k$ and $\omega'_k$ share the same slope in the long-wavelength limit $k \to 0$, ensuring that the paramagnon velocity remains unchanged at small momentum. This reflects the hydrodynamic nature of the long-wavelength spin response, which should not depend on the details of the lattice reconstruction. The dynamical character of the CDF is encoded in a single parameter $\Gamma$, which captures the precursor effects of the lattice reconstruction through the phenomenological self-energy
\[
\widetilde{\Sigma}_k = 
\frac{(\omega'_k-\omega_k)^{2}}{(\omega'_k-\omega_k)^{2} + \Gamma^{2}}\left(\omega'_k-\omega_k-\mathrm i\Gamma\right).
\]
The corresponding renormalized paramagnon dispersion reads
\[
\omega = \widetilde{\omega}_k \equiv \omega_k +\mathrm{Re}\,\widetilde{\Sigma}_k=\omega_k+\frac{(\omega'_k - \omega_k)^{3}}
{(\omega'_k - \omega_k)^{2} + \Gamma^{2}},
\]
which interpolates between $\omega_k$ and $\omega'_k$ in the limits $\Gamma \to \infty$ (no reconstruction) and $\Gamma \to 0$ 
(static lattice reconstruction), respectively. The associated spectral broadening is given by
\[
\widetilde{\Gamma}_k \equiv -\mathrm{Im}\,\widetilde{\Sigma}_k=\frac{(\omega'_k - \omega_k)^{2}\Gamma}
{(\omega'_k - \omega_k)^{2} + \Gamma^{2}},
\]
which vanishes both for $\Gamma \to \infty$ and $\Gamma \to 0$, as well as in the long-wavelength limit $k \to 0$, ensuring that the low-energy paramagnon spectrum is unaffected by the precursor fluctuations.
As shown in Fig.~\ref{fig:figure5}, this phenomenological self-energy captures the softening at $q_{\mathrm{CDF}}$, while also predicting a renormalization at the zone boundary which could be removed by introducing a momentum-dependent $\Gamma$. The phenomenological parameter $\Gamma$ controls both the magnitude of the softening and the associated spectral broadening, and should be regarded as an effective measure of the dynamical and incoherent character of the CDF, which modify the paramagnon spectrum without establishing long-range and static lattice reconstruction. At the same time, $\Gamma$ also reflects the strength of the electron--phonon coupling, with stronger coupling corresponding to smaller $\Gamma$ and hence to a more pronounced softening.

The emerging picture is therefore the following. In Hg1223---the cuprate with the highest $T_{\mathrm{c}}$---the various excitations that contribute to the ground state appear maximally intertwined. Here, the EPC is exceptionally strong, as suggested by both theory and experiment. 
In turn, this facilitates the formation of particularly robust CDF, which can then act as mediators between charge and spin degrees of freedom, giving rise to a sizable spin--phonon coupling.


Our measurements provide direct experimental evidence for such coupling, reinforcing earlier indirect signatures inferred from the observation of phonon chirality in thermal Hall conductivity experiments~\cite{grissonnanche2020chiral}. This spin–phonon coupling, here amplified but potentially universal across the cuprate family, has been theoretically proposed as a key ingredient of unconventional superconductivity~\cite{gao2023chiral}. In this context, recent theoretical work further supports a scenario in which the EPC is enhanced by magnetic correlations via chiral phonons, providing a natural route to boost pairing interactions~\cite{wang2026rotational}. Remarkably, analogous effects have been demonstrated in multiferroic materials, where spin–phonon coupling correlates with the emergence of superconductivity~\cite{ideue2017giant, weber2022emerging}. 

Our results may also provide a new perspective on the earlier observation of paramagnon softening in LBCO. At first sight, that system appears difficult to reconcile with the present picture, since the softening occurs in a regime dominated by static stripe order and strongly suppressed superconductivity. However, recent experiments have shown that suppressing static stripes by in-plane strain dramatically enhances superconductivity while preserving, and possibly promoting, their dynamical counterpart~\cite{sazgari2026stripe}. These findings suggest that the ingredient relevant for pairing is not the static charge order itself, but rather the associated dynamical charge--spin correlations.
Within this framework, the paramagnon softening observed in LBCO and Hg1223 may share a common origin in dynamical charge--spin fluctuations. The key difference is that, in LBCO at ambient conditions, these fluctuations coexist with a static stripe component that competes with superconductivity, whereas in Hg1223 the dynamical component emerges without developing long-range static order. This distinction naturally reconciles the occurrence of paramagnon softening in both materials despite their significantly different superconducting properties.

Altogether, these findings point to a paradigm in which high-$T_c$ superconductivity emerges from the cooperative interplay of lattice, charge, and spin fluctuations.

\section{\label{sec:methods}Methods}
\subsection*{Samples}\label{samples}
The single crystal of HgBa$_2$Ca$_2$Cu$_3$O$_{8+\delta}$ (Hg1223) used for this study was grown by a self-flux method, as reported in Ref. \cite{loret2017crystal}. The resulting crystals  exhibit excellent surface quality suitable for spectroscopic investigations, and their superconducting transition shows a narrow broadening of $\sim$2 K, indicating high homogeneity of the superconducting properties. Consistently, the hole doping can be systematically tuned by appropriate heat treatments, which for the first time to our knowledge allowed us to access even strongly underdoped regimes (see Fig.~\ref{fig:figure1}b). For the RIXS experiment we selected a crystal with a superconducting transition temperature $T_{\mathrm{c}}=112$ K, measured by magnetic susceptibility. This value corresponds to a doping level $p\approx0.12$, estimated both from the empirical parabolic $T_{\mathrm{c}}(p)$ dependence \cite{presland1991general} together with the knowledge of $T_{\mathrm{c,max}}=135$ K, critical temperature of optimally doped Hg1223  (violet circles in Fig.~\ref{fig:figure1}b) and from the method established in Ref. \cite{liang2006evaluation}, which combines the measured $T_{\mathrm{c}}$ with the $c$-axis lattice parameter determined by X-ray diffraction (green circles in Fig.~\ref{fig:figure1}b). Prior to RIXS, the crystal surface was polished in three successive steps using diamond pastes with grain sizes of 3~$\mu$m, 1~$\mu$m and 0.1~$\mu$m. This procedure was used to obtain a smooth surface with minimal submicron scratches and to remove possible surface oxidation, which may have developed after previous measurements or repeated thermal cycles. X-ray absorption measurements were then performed to set the incident energy at the Cu $L_3$ edge. The spectrum shows, besides the main peak, a weak shoulder $\sim$2 eV higher in energy, attributed to $\left| d^{10}\underline{L} \right\rangle$ ligand-hole states. Its reduced intensity, compared to the strong peak expected at optimal doping, confirms that the sample is underdoped. During measurements, the sample demonstrated spatial homogeneity, as identical spectra were obtained from different regions under identical conditions, and stability under the X-ray beam, with no measurable changes over time.

To compare the energy profile of charge-density fluctuations (CDF) in Hg1223 with another cuprate family, we investigated two YBa$_2$Cu$3$O$_{7-\delta}$ (YBCO) thin films of thickness $t=100$ nm, grown by pulsed laser deposition on (001)-oriented SrTiO$_3$ substrates with lateral dimensions of 5$\times$5 mm$^2$ (see growth details in Ref.~\cite{ArpaiaGrowth}). The films exhibited zero-resistance critical temperatures of 65 K and 86 K, tuned by post-growth oxygen annealing at 1.2$\times$10$^{-2}$ Torr and 6.5$\times$10$^{2}$ Torr, respectively. The corresponding hole concentrations, $p=0.13$ (underdoped) and $p=0.19$ (slightly overdoped), were estimated following Ref. \cite{liang2006evaluation}.

\subsection*{Fit of RIXS spectra}\label{analysis}
Prior to the analysis, the RIXS spectra were corrected for self-absorption and normalized to the integral of the inter-orbital  $dd$ excitations in the energy-loss range [1–3 eV], enabling direct comparison across different experimental conditions. The self-absorption correction was performed individually for each momentum transfer, accounting for sample thickness, scattering geometry, and the energy and polarization dependence of the absorption coefficients, as obtained from XAS measurements \cite{NatCommRArpaia}.

The fitting procedures for ultrahigh-resolution (UHR) and medium-resolution (MR) spectra differ due to the inability to resolve the different low energy features in the MR case. 
For UHR spectra, the intensity below 1~eV was decomposed into six components: (i) a resolution-limited elastic peak, (ii) a low-energy peak associated with CDF, (iii) a BS phonon, (iv) its overtone (with the energy constrained to twice that of the BS phonon), (v) a paramagnon contribution, and (vi) an electron–hole (e–h) continuum background. The quasi-elastic components were modelled by Gaussian functions reflecting the instrumental resolution. The paramagnon intensity was modelled using a damped harmonic oscillator (DHO) form for the magnetic susceptibility $\chi''(Q,\omega)$. 
Since the RIXS scattering cross section is proportional to the spin dynamical structure factor $S(Q,\omega)$, and $S(Q,\omega)\propto \chi''(Q,\omega)$~\cite{haverkort2010theory, PengPRB}, the paramagnon intensity was modelled as
\begin{equation}
    I(\omega) \propto \frac{\gamma \omega}{\left( \omega^{2}-\omega_{0}^{2}\right)^{2}+4\gamma^{2}\omega^{2}},
\end{equation}
where $\omega_{0}$ and $\gamma$ are the undamped frequency and damping factor according to the DHO model. 
The e–h continuum was approximated by a linear background (which also takes into account the tails of $dd$ excitations at high energies). 
For MR spectra, the quasi-elastic region was reduced to two Gaussian components (elastic + charge order, and phonon), while the higher-energy contributions were treated identically to the UHR case. 

\subsection*{Extraction of $J^{\mathrm{eff}}$ using Linear Spin-wave Theory}

To establish the baseline of the paramagnon dispersion fitting, we first considered a simplified 
Heisenberg model based on nearest-neighbour (nn) in-plane ($J_{\parallel}$) and out-of-plane ($J_{\perp}$) 
exchange interactions within the unit cell. For simplicity, $J_{\perp}$ is assumed to be the same 
between the inner and the two outer CuO$_2$ planes. The Hamiltonian is described as
\begin{equation}
    \mathcal{H} = J_{\parallel} \sum_{\langle i,j \rangle, \alpha} \mathbf{S}_{i,\alpha} \cdot \mathbf{S}_{j,\alpha} 
    + J_{\perp} \sum_{i} \left( \mathbf{S}_{i,1} \cdot \mathbf{S}_{j,2} + \mathbf{S}_{i,2} \cdot \mathbf{S}_{j,3} \right),
\end{equation}
where $\alpha = 1,2,3$ labels the three CuO$_2$ planes, with $\alpha = 2$ corresponding to the inner plane 
and $\alpha = 1,3$ to the outer planes. The index $j$ runs over lattice sites within a given plane, and 
$\langle i,j \rangle$ denotes nearest-neighbour pairs within the same plane. 

However, the nn Heisenberg model fails to capture the paramagnon dispersion along both $(H,0)$ and $(H,H)$ 
directions, as shown by the dashed lines in Fig.~\ref{fig:figure2}(f)-(g). To accurately describe the dispersion, we employed a 
linear spin-wave framework incorporating higher-order in-plane interactions, in which the Heisenberg 
Hamiltonian is typically expressed as~\cite{roger1989cyclic, coldea2001spin}:
\begin{equation}
\begin{split}
    \mathcal{H} &= J \sum_{\langle i,j \rangle} \mathbf{S}_{i} \cdot \mathbf{S}_{j} 
    + J^{\prime} \sum_{\langle i,i' \rangle} \mathbf{S}_{i} \cdot \mathbf{S}_{i'} 
    + J^{\prime\prime} \sum_{\langle i,i'' \rangle} \mathbf{S}_{i} \cdot \mathbf{S}_{i''} \\
    & \quad + J_{c} \sum_{\langle i,j,k,l \rangle} \Big[ 
    (\mathbf{S}_{i} \cdot \mathbf{S}_{j})(\mathbf{S}_{k} \cdot \mathbf{S}_{l}) 
    + (\mathbf{S}_{i} \cdot \mathbf{S}_{l})(\mathbf{S}_{k} \cdot \mathbf{S}_{j}) 
    - (\mathbf{S}_{i} \cdot \mathbf{S}_{k})(\mathbf{S}_{j} \cdot \mathbf{S}_{l}) \Big],
\end{split}
\end{equation}

where $J$, $J'$ and $J''$ denote the first-, second- and third-nearest-neighbour super-exchange integrals, 
and $J_{c}$ is the ring (cyclic) exchange interaction within a Cu plaquette. 
The interlayer coupling $J_{\perp}$ can also be included, leading to the effective Hamiltonian~\cite{delannoy2009low, PengInfluence}:
\begin{equation}
    \mathcal{H} = J_{1}^{\mathrm{eff}} \sum_{\langle i,j \rangle} \mathbf{S}_{i} \cdot \mathbf{S}_{j} 
    + J_{2}^{\mathrm{eff}} \sum_{\langle i,i' \rangle} \mathbf{S}_{i} \cdot \mathbf{S}_{i'} 
    + J_{3}^{\mathrm{eff}} \sum_{\langle i,i'' \rangle} \mathbf{S}_{i} \cdot \mathbf{S}_{i''} 
    + J_{\perp} \sum_{\langle i,j' \rangle} \mathbf{S}_{i} \cdot \mathbf{S}_{j'},
\end{equation}
with the effective exchange interaction 
$J_{1}^{\mathrm{eff}} = J - J_{c}/2$, 
$J_{2}^{\mathrm{eff}} = J' - J_{c}/4$, 
and $J_{3}^{\mathrm{eff}} = J''$. 

In our analysis, we additionally included a fourth-nearest-neighbour term $J_{4}^{\mathrm{eff}}$, which provides a more accurate description of the paramagnon dispersion across different cuprate compounds without introducing unnecessary free parameters. 
This extended framework will be referred to as the $4J^{\mathrm{eff}}$ model.  

The fits were performed using \textsc{SpinW}~\cite{toth2015linear}, a MATLAB-based library developed at PSI for simulating spin-wave spectra within linear spin-wave theory. 
Following Ref.~\cite{PengInfluence}, the $4J^{\mathrm{eff}}$ model was constrained by fixing $J_{1}^{\mathrm{eff}}$ to the value obtained independently from (Eq.~4), thereby reducing the number of free fitting parameters.

\subsection*{Analysis of momentum-dependent integrated intensities}\label{Greven}

The integrated RIXS intensity was determined as a function of momentum in each of the 60~meV wide intervals into which the spectra at 110~K and 300~K were divided (see Fig. \ref{fig:figure4}). 
For each interval, the momentum dependence was analyzed by fitting with a single Gaussian function superimposed on a second-order polynomial background, both treated as free parameters. 
The Gaussian peak position was constrained to vary within $q=0.20–0.40$. 

In the energy range overlapping with the BS phonon regime, the phonon contribution was explicitly modelled using the experimentally determined $q$-dependence of the phonon intensity extracted from the fits in Fig.~\ref{fig:figure2}, which follows the expected $\sin^{2}(\pi q)$ form with an enhancement due to coupling to CDF. 
The polynomial background was chosen to reproduce the overall concave envelope of the spectra, while excluding contributions from both the CDF peak at $q_{\mathrm{CDF}}$ and the phonon-related signal.  

We believe that the subtraction of the BS phonon contribution is the key difference between our results and those reported for Hg1201 in Ref.~\cite{yu2020unusual}, where in principle a similar data analysis was performed. In that work, the larger integrated intensity observed at high momentum values—originating from BS phonons and their overtones in the 60–140~meV range—leads to an underestimation of the CDF peak height at $q_{\mathrm{CDF}}$. As a consequence, while in our case the CDF intensity gradually decreases with increasing energy, exhibiting at low temperature a broad tail extending up to 300~meV, in Ref.~\cite{yu2020unusual} the lack of phonon subtraction likely explains the abrupt drop of the integrated intensity in the intermediate energy range and the apparent splitting of the CDF signal into two components, one at low and the other at high energy.

\subsection*{Energy profile of CDF in YBCO}\label{YBCO}

To determine the energy profile of CDF in a cuprate HTS compound that does not display paramagnon softening, we focused on two YBa$_2$Cu$_3$O$_{7-\delta}$ (YBCO) thin films with hole doping levels $p=0.13$ and $p=0.19$. 

RIXS measurements at the Cu $L_3$ edge ($\approx 931$~eV) were performed at the ID32 beamline of the ESRF, with a combined energy resolution of $\sim$40~meV and $\sigma$-polarized incident x-rays. The scattering angle was fixed at $2\theta = 149.5^\circ$. Spectra were acquired across the momentum range 0.14–0.48~r.l.u. along the $(H,0)$ direction. The underdoped film ($p=0.13$) was measured at 20~K, while the slightly overdoped film ($p=0.19$) was measured at 80~K. 

After acquisition, the spectra were corrected for self-absorption and normalized to the integral of the inter-orbital $dd$ excitations in the [1–3~eV] energy-loss range. The subsequent analysis followed the same procedure as for Hg1223 (see Fig.~\ref{fig:figure4}), including the choice of background. For the $p=0.13$ film, the elastic signal is dominated by CDW. To isolate the CDF component from the $q$-dependence of the quasi-elastic intensity in the 60~meV intervals at low temperature, the peak at $q \approx q_{\mathrm{CDF}}$ was fitted with two Gaussian functions: a narrower one associated with CDW and a broader one associated with CDF (see Supplementary Fig. \ref{fig:YBCOUD}). Although the strong CDW intensity in the quasi-elastic region introduces some uncertainty in the determination of the CDF amplitude, a robust outcome of the analysis is that above 80~meV there is no enhancement of the integrated RIXS intensity around $q_{\mathrm{CDF}}$. This implies that the high-energy CDF component—the broad tail identified in Hg1223—is absent in YBCO (see Supplementary Fig. \ref{fig:CDFYBCO}).

\subsection*{Phenomenological theory of the paramagnon softening}\label{th_pheno}

Aiming at a theoretical description of the paramagnon softening observed in our experiments, we argue that, 
for a sufficiently strong electron-phonon coupling, a static CDW with momentum $q_{\mathrm{CDW}}$ should induce a full 
reconstruction of the electron bands: states at $k$ are coherently coupled to states at $k\pm q_{\mathrm{CDW}}$, 
leading to a folding of the band structure into a reduced Brillouin zone and to the opening of CDW gaps at the new zone boundaries. 
Since paramagnons correspond microscopically to spin-flip particle-hole excitations built on top of the underlying electron structure, 
a folded and gapped electron dispersion can induce a corresponding folding of the paramagnon dispersion. The multiple branches that 
emerge in this situation do not represent independent collective modes, rather distinct spin-flip particle-hole excitations between 
the reconstructed bands. However, our measurements show that the paramagnon softening should instead be attributed to CDF, and is indeed observed even in the absence of CDW. On the one hand, in close analogy to a dense liquid, which can be regarded as a `failed' solid, CDF may be viewed as `failed' CDW: they retain the same underlying tendency towards charge ordering, but remain dynamic and strongly coupled to the lattice. On the other hand, even when CDW are detected in RIXS experiments, their spectral weight is weak, reflecting the sparse and fragile nature of the corresponding CDW puddles, as well as their much weaker coupling to the lattice. This picture is consistent with our recent results in Ref.~\cite{fedele2026electron}, where the partial phonon softening was shown to be driven by CDF rather than by CDW. These circumstances may explain why CDW play a marginal role (if any) in determining the paramagnon
softening.
To describe CDF as a `failed' CDW, we start considering a static lattice reconstruction and denote by 
$\omega_k$ and $\omega'_k$ the paramagnon dispersion in the original and reconstructed state, with lattice
spacing $a$ and $a'>a$, respectively. 
We impose that $\omega_k$ and $\omega'_k$ share the same slope in the long-wavelength limit 
$k \to 0$, ensuring that the paramagnon velocity remains unaffected at small momentum. This reflects the hydrodynamic 
nature of the long-wavelength spin response, which should not depend on the details of the lattice 
reconstruction. We can formally 
represent the change in the paramagnon Green's function produced by the lattice reconstruction,
\begin{equation}
D(k,\omega) = (\omega - \omega_k)^{-1}
\;\longrightarrow\;
D'(k,\omega) = (\omega - \omega'_k)^{-1}
\equiv
(\omega - \omega_k - \Sigma_k)^{-1},
\end{equation}
in terms of a static self-energy $\Sigma_k \equiv \omega'_k - \omega_k$, that 
is understood as a phenomenological quantity encoding the modification of the dispersion induced by a fully developed 
lattice reconstruction, rather than as a microscopically derived quantity.
To describe the precursor effects of the full lattice reconstruction, one can introduce a broadened version of 
the self-energy, e.g.,
\begin{equation}
\widetilde{\Sigma}_k \equiv
\frac{\Sigma_k}{\Sigma_k + i\Gamma}\,\Sigma_k,
\end{equation}
which interpolates continuously between $\Sigma_k$ (fully reconstructed state) and $0$ (absence of reconstruction), for $\Gamma \to 0$ 
and $\Gamma \to \infty$, respectively. For simplicity, $\Gamma$ will be assumed to 
be momentum independent.
{The broadened self-energy acquires a real and an imaginary part,
\begin{equation}
\widetilde{\Sigma}_k = \widetilde{\Sigma}'_k + \mathrm i\widetilde{\Sigma}''_k
\equiv
\frac{\Sigma_k^{3}}{\Sigma_k^{2} + \Gamma^{2}}
- \mathrm i \frac{\Sigma_k^{2}\Gamma}{\Sigma_k^{2} + \Gamma^{2}},
\end{equation}
the latter encoding the finite lifetime associated with the fluctuating precursor regime.
{The pole of the paramagnon Green's function is located at
\begin{equation}
\omega = \widetilde{\omega}_k \equiv\omega_k + \widetilde{\Sigma}'_k
= \omega_k +\frac{(\omega'_k - \omega_k)^{3}}
{(\omega'_k - \omega_k)^{2} + \Gamma^{2}},
\end{equation}
which interpolates between $\omega_k$ and $\omega'_k$ in the limits $\Gamma \to \infty$ and $\Gamma \to 0$, respectively.
The corresponding spectral broadening is
\begin{equation}
\widetilde{\Gamma}_k \equiv -\widetilde\Sigma''_k =
\frac{\Sigma_k^{2}\Gamma}{\Sigma_k^{2} + \Gamma^{2}}=\frac{(\omega'_k - \omega_k)^{2}\Gamma}
{(\omega'_k - \omega_k)^{2} + \Gamma^{2}},
\end{equation}
which vanishes both for $\Gamma \to 0$ (static lattice reconstruction) and $\Gamma \to \infty$ (no reconstruction). Importantly, 
$\widetilde{\Gamma}_k$ also vanishes for $k \to 0$, ensuring that the long-wavelength paramagnon spectrum remains unaffected by the 
precursor fluctuations. 
For the purpose of illustration, we consider a one-dimensional case with $a' = 3a$, and take
\begin{equation}
\omega_k = \left|\sin\!\left(\tfrac{1}{2}ka\right)\right|,
\qquad
\omega'_k = \tfrac{1}{3}
\left|\sin\!\left(\tfrac{3}{2}ka\right)\right|,
\end{equation}\label{eq:paradispersions}
where the paramagnon velocity is set equal to one. The resulting softened paramagnon spectrum, in the case $\Gamma=2$, is illustrated
in Fig.~\ref{fig:figure5} of the main text.

\subsection*{Microscopic theory of the paramagnon excitations in a CDW system}\label{sec:meth:micr}
To cast the phenomenological description adopted in the previous section on a microscopic footing,
 we can consider a 2D electronic system subject to CDW scattering, described by the Hamiltonian
  \begin{equation}
    H=\sum_{{\bf k},\sigma}\varepsilon_{\bf k} c_{{\bf k},\sigma}^\dagger c_{{\bf k},\sigma}
    +\Delta \sum_{k,\sigma}\left\lbrack c_{{\bf k}+{\bf Q}_c,\sigma}^\dagger c_{{\bf k},\sigma}
    + c_{{\bf k},\sigma}^\dagger c_{{\bf k}+{\bf Q}_c,\sigma}\right\rbrack
  \end{equation}
  with dispersion $\varepsilon_{\bf k}=-2t[\cos(k_xa)+\cos(k_ya)]-4t'\cos(k_xa)\cos(k_ya)$.
  The idea is to explore the evolution of the paramagnon dispersion adding to this system a strongly interacting and dynamical CDF. 
  
 In the following, the lattice constant is set to $a=1$ and the nearest-neighbor hopping is fixed to $t'/t=-0.2$. Energies are expressed in units of the nearest-neighbor hopping $t$. The CDW scattering momentum is chosen as ${\bf q}_c=\tfrac{2\pi}{3a} (1,0)$, close to the value of $q_\mathrm{CDF}$ observed in the RIXS spectra. The Hamiltonian can then be diagonalized in the reduced Brillouin zone $-\tfrac{\pi}{3} \le k_x \le \tfrac{\pi}{3}$. The transformation $c_{{\bf k}+n{\bf Q}_c,\sigma}=\sum_{p=1}^3 \Phi_{\bf k}(n,p)\, a_{{\bf k},\sigma}(p)$ ($n=1\dots 3$) yields three bands $E_{\bf k}$ in the reduced zone, and can be used to calculate the (bare) spin-spin correlation function in the CDW state,
    \begin{equation}\label{eq:chi}
  \chi^{+-,0}_{n,m}({\bf q},\omega)=\mathrm i\frac{Z^2}{N}\int\,\mathrm dt\,\mathrm e^{-\mathrm i\omega t}\,\langle{\cal T} 
  S^+_{{\bf q}+n{\bf Q}_c}(t) S^-_{-{\bf q}-m{\bf Q}_c}(0)\rangle
    \end{equation}
where $Z$ is a renormalization factor accounting for fluctuation effects and ${\cal T}$ is the time-ordering operator. In contrast to the homogeneous system, $\chi^{+-,0}_{n,m}({\bf q},\omega)$ can transfer multiples of the CDW vector between spin-flip excitations for $n \ne m$. To mimic the case of CDF, i.e., in the absence of symmetry breaking, we neglect these processes and retain only the diagonal components  $\chi^{+-,0}_{n,n}(q,\omega)$. Including local correlations within a Hubbard interaction $\sim U$, the paramagnon excitations are obtained via a random-phase approximation resummation~\cite{markiewicz2007paramagnon}
    \begin{equation}
      \chi^{+-}_{n,n}({\bf q},\omega)=\frac{\chi^{+-,0}_{n,n}({\bf q},\omega)}{1-U\chi^{+-,0}_{n,n}({\bf q},\omega)} \; ,
    \end{equation}
where we set $U/t=4$. Supplementary Fig.~\ref{fig1M}(a) shows the imaginary part of $\chi^{+-}_{n,n}({\bf q},\omega)$ as a function of momentum and energy for a CDW parameter $\Delta/t=0.2$. To account for the width of the CDF excitations, we further convolve  $\chi^{+-}_{n,n}({\bf q},\omega)$ with a Lorentzian of width $0.3\,t$,  yielding the spectra shown in Supplementary Fig.~\ref{fig1M}(b). For typical cuprate parameters $t\approx 250$~meV~\cite{markiewicz2005one}, this corresponds to a broadening of $\sim 80$~meV, consistent with the intrinsic FWHM of the CDF peak.

\section{\label{sec: ack}Acknowledgments}

We are grateful for insightful discussions with Emily Zhang and Tom Devereaux. 
The experimental data were collected at the beam line ID32 of the ESRF during the experiments HC4150 and HC4825. The YBCO thin films were grown at Myfab Chalmers. K.R. acknowledges the Knut and Alice Wallenberg Foundation via the Wallenberg Initiative Materials Science for Sustainability WISE. S.C. acknowledges support from the University of Rome Sapienza, under the projects Ateneo 2023 (RM123188E830D258), Ateneo 2024 (RM124190C54BE48D), and Ateneo 2025 (RP125199B9FDBFE4). A.S. and D.C. acknowledge support from the ANR grant NEPTUN (ANR-19-CE30-0019-01).

\section{\label{sec: author}Author Contributions}

R.A. conceived and designed the experiment with suggestions from G.G.. L.M., R.S., N.B.B., R.A., M.M.S. and G.G. performed the RIXS measurements. D.C. and A.S. grew and characterized the Hg1223 single crystals. K.R. and R.A. analysed the RIXS experimental data. G.S. and S.C. developed the theoretical modelling. R.A., K.R., M.F., G.M., A.S., G.G. and M.M.S. discussed and interpreted the results. R.A., M.F. and K.R. wrote the manuscript with contributions from all authors.

\section{\label{sec: info}Additional Information}

Correspondence and requests for materials should be addressed to Giacomo Ghiringhelli and Riccardo Arpaia.  

\section{\label{sec:figures}Figures}
\begin{figure}[H]
    \centering
    \includegraphics[width=1.02\linewidth]{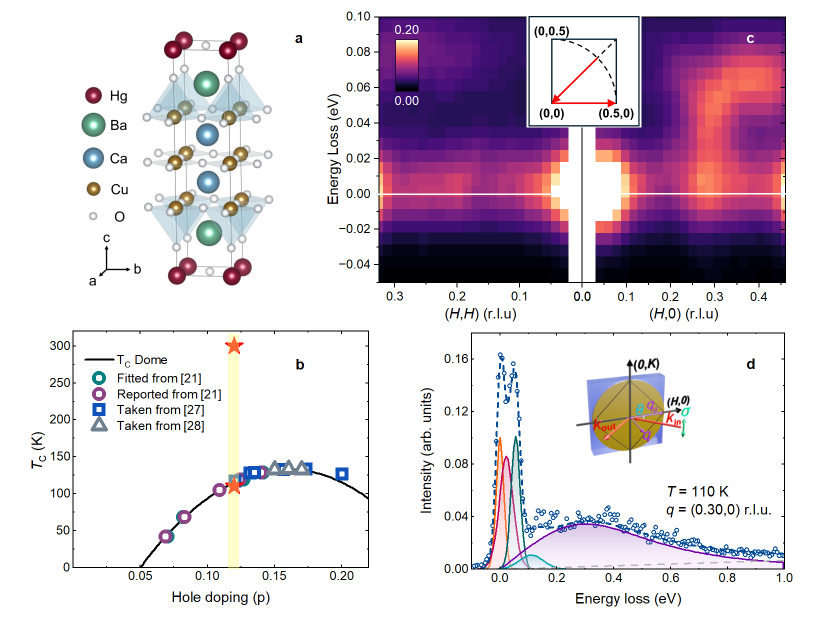}
    \caption{\textbf{Crystal structure, phase diagram, and experimental conditions of HgBa$_2$Ca$_2$Cu$_3$O$_{8+\delta}$ (Hg1223).} \textbf{(a)} Crystal structure of Hg1223. \textbf{(b)} Superconducting dome of Hg1223, with $T_{\mathrm{c}}$ values from samples both as those of our study and from the literature. Their dependence follows a parabolic relationship, without any hint of anomalous $T_{\mathrm{c}}$ suppression in the underdoped region. RIXS spectra were acquired in medium resolution mode at several temperatures between 20 and 300 K (yellow bar), and in high resolution mode at 110 K, i.e., just below the superconducting transition, and at room temperature (stars). \textbf{(c)} High-resolution ($\Delta E = 32$ meV) RIXS maps measured at $T=110$ K along the $(H,0)$ and $(H,H)$ directions, corresponding to the red path M–$\Gamma$–X sketched in the central inset.
    \textbf{(d)} Example of fit at $q=(0.3,0)$ r.l.u.. The orange, magenta, green, and light green Gaussians, together with the violet area and the gray dashed line, represent respectively the elastic peak (dominated by the specular contribution visible at $\Gamma$ in panel (c)), charge density fluctuations (CDF), bond-stretching phonons, bond-stretching overtone, paramagnons, and the weakly energy-dependent background from the electron–hole continuum. Additional details on the fitting procedure are reported in the Methods section. In the inset, the experimental geometry is included with $\sigma$ incident polarization, chosen to enhance the charge response and maximize its impact on the magnetic excitations.}   
    \label{fig:figure1}
\end{figure}

\begin{figure}[H]
    \centering
    \includegraphics[width=1.01\linewidth]{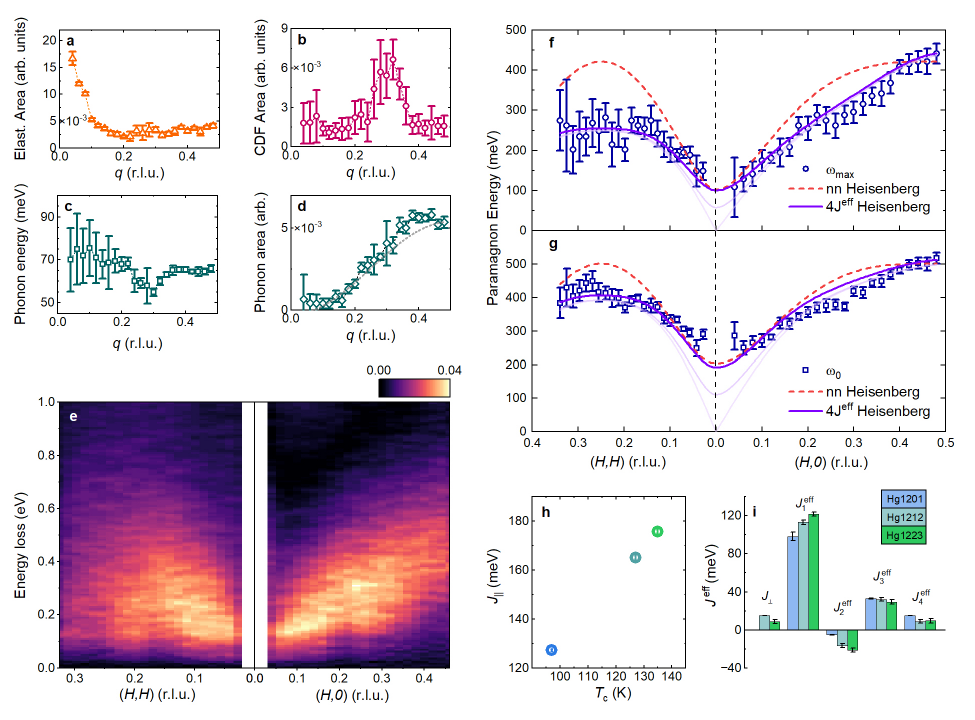}
    \caption{\textbf{Charge, lattice, and magnetic excitations in Hg1223 from RIXS.} From the fits as a function of momentum along the $(H,0)$ direction we extract: \textbf{(a)} the area of the elastic line, \textbf{(b)} the area of the CDF peak, \textbf{(c)} the dispersion and \textbf{(d)} the area of the bond-stretching phonons. \textbf{(e)} Inelastic RIXS maps along the $(H,0)$ and $(H,H)$ directions obtained by subtracting from the raw spectra all low-energy contributions identified in the fits of panels (a–d), so as to single out the spectral weight associated with the paramagnons. \textbf{(f)} Dispersion of the paramagnon energy $\omega_{\mathrm{max}}$, fitted using both the nearest-neighbour Heisenberg model (dashed line) and a phenomenological linear spin-wave model with four nearest-neighbour coupling parameters. \textbf{(g)} Same analysis as in panel (f) for the paramagnon energy $\omega_{0}$. \textbf{(h)}  In-plane exchange interaction $J_{||}$ obtained from the nearest-neighbour Heisenberg model as a function of the maximum $T_{\mathrm{c}}$ achievable for Hg1223, $T_{\mathrm{c}}^{\mathrm{max}}$, compared with the values extracted for the single- and double-layer members of the Hg family (data from Ref.~\cite{wang2022paramagnons}). \textbf{(i)}  Effective parameters of the phenomenological spin-wave model including four in-plane couplings $J_{\mathrm{eff}}$ and the interlayer exchange $J_\perp$, compared across Hg1201, Hg1212 and Hg1223.}   
    \label{fig:figure2}
\end{figure}

\begin{figure}[H]
    \centering
    \includegraphics[width=1.05\linewidth]{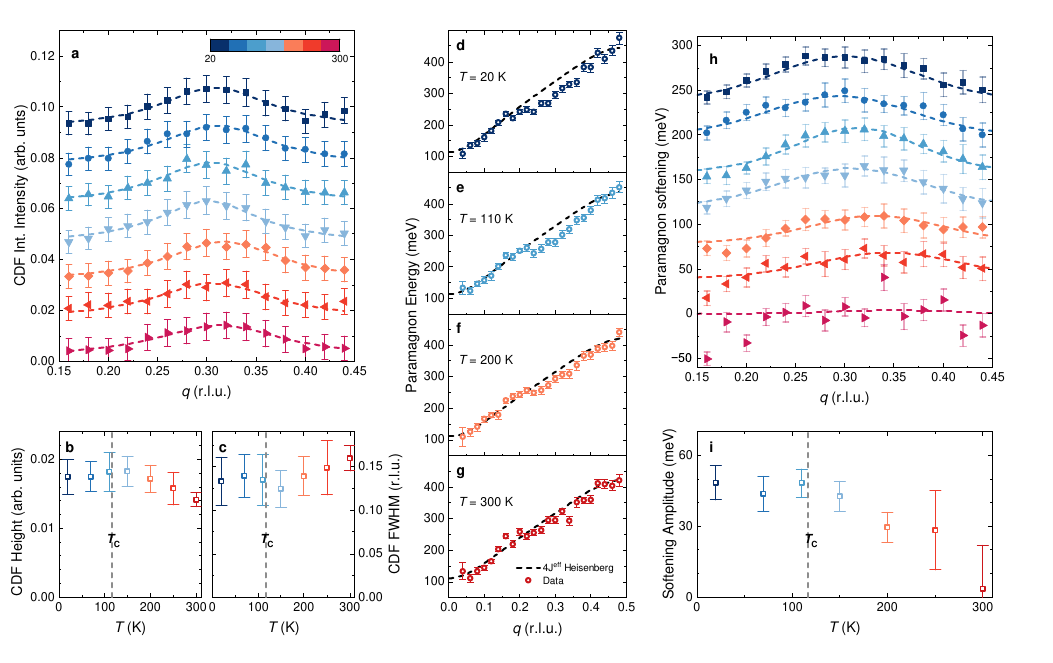}
    \caption{\textbf{Interplay of paramagnon softening and charge-density fluctuations.} (\textbf{a}) The CDF peak at different temperatures between 20 and 300 K, highlighted by the RIXS intensity integrated in the range [–0.1, 0.035] eV.  (\textbf{b,c}) Height and FWHM of the Lorentzian profiles used to fit the CDF peak. Both intensity and correlation length increase upon cooling and saturate below $T_{\mathrm{c}}$. (\textbf{d–g}) Paramagnon energy $\omega_{\mathrm{max}}$ extracted from the fits at various temperatures, compared with the phenomenological linear spin-wave model including four nearest-neighbour exchange parameters.  (\textbf{h}) Paramagnon softening at each temperature, quantified by the deviation between experimental data and the fitted spin-wave dispersion of panels (d–g). The resulting difference exhibits a peak, which can be fitted by a Lorentzian profile, centered at a momentum very close to that of the CDF.  (\textbf{i}) Temperature dependence of the paramagnon softening, which closely mirrors the behaviour of the CDF, with an increase upon cooling and a plateau in the superconducting state.
    }   
    \label{fig:figure3}
\end{figure}

\begin{figure}[H]
    \centering
    \includegraphics[width=1\linewidth]{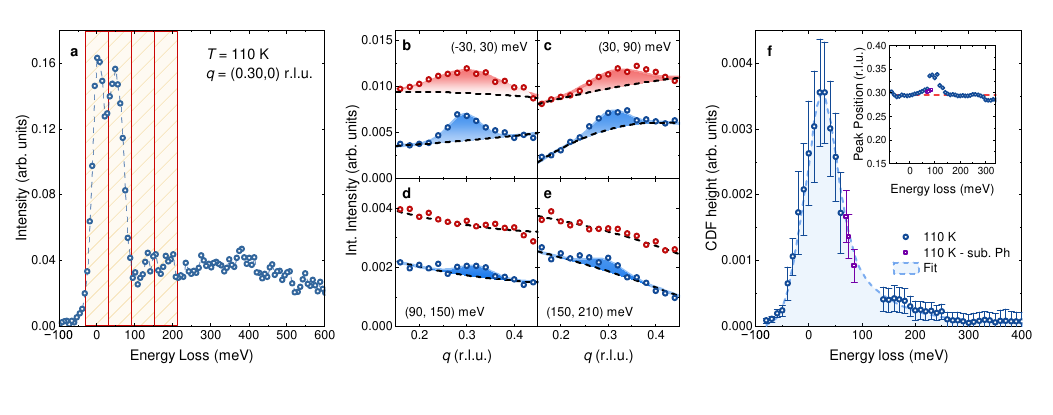}
    \caption{\textbf{Broad energy range of charge-density fluctuations overlapping with paramagnons.} (\textbf{a}) To assess the energy extent of the CDF peak, each RIXS spectrum was divided into several energy intervals, and the integrated intensity was extracted for each of them.  (\textbf{b–e}) Example of this procedure at $T=110$ K and 300 K, using 60 meV–wide adjacent intervals covering the range from –30 to 210 meV. While the CDF peak is maximal in the quasi-elastic region [–30, 30] meV, at low temperature it survives up to 210 meV, whereas at room temperature this high-energy component vanishes.  (\textbf{f}) To better visualize the evolution of the CDF peak with energy, at $T=110$ K the analysis was repeated with 60 meV–wide intervals shifted by only 10 meV. As highlighted in panels (b-e) by the colored region, for each interval a peak centered around $q\approx0.30$ r.l.u. is identified, and its height plotted as a function of energy. This reveals a broad and weak tail of the CDF extending up to $\sim$300 meV, where the paramagnons reside. In the 70–120 meV range the signal is dominated by bond-stretching phonons; after subtraction of this contribution, the residual CDF component is highlighted (violet points). 
    }   
    \label{fig:figure4}
\end{figure}

\begin{figure}[H]
    \centering
    \includegraphics[width=1\linewidth]{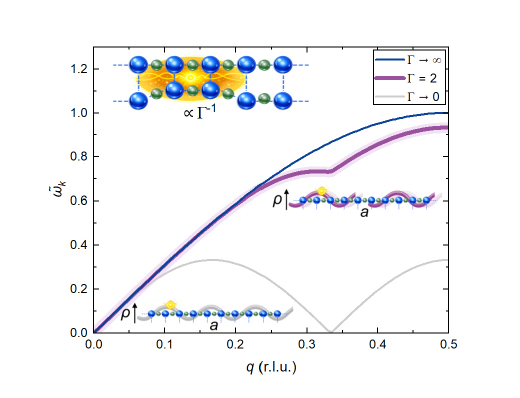}
    \caption{\textbf{Phenomenological model for paramagnon softening induced by charge-density fluctuations.} 
    Renormalized paramagnon dispersion $\widetilde{\omega}_k$ obtained from the phenomenological self-energy describing the coupling between paramagnons and CDF.
    The bare dispersion $\omega_k$ ($\Gamma \to \infty$, blue line) evolves towards the reconstructed dispersion $\omega'_k$ ($\Gamma \to 0$, grey line) through an intermediate regime ($\Gamma=2$, violet line), where a pronounced softening develops at $q_{\mathrm{CDF}}$. 
    The parameter $\Gamma$ captures the dynamical character of the CDF and effectively encodes the strength of the electron--phonon coupling: smaller $\Gamma$ corresponds to slower, more correlated fluctuations and leads to stronger softening. 
    The upper-left schematic illustrates the inverse relation between EPC and $\Gamma$, whereas the lower schematics associate the partial and complete softening of the paramagnon dispersion with dynamical (violet) and quasi-static (grey) electronic modulations, respectively.}
    \label{fig:figure5}
\end{figure}

\section{Supplementary Figures}

\newenvironment{myextfigure}[1][]%
  {\captionsetup{type=extfigure}\begin{figure}[#1]}%
  {\end{figure}}

\begin{extfigure}[H]
    \centering
    \includegraphics[width=1.1\linewidth]{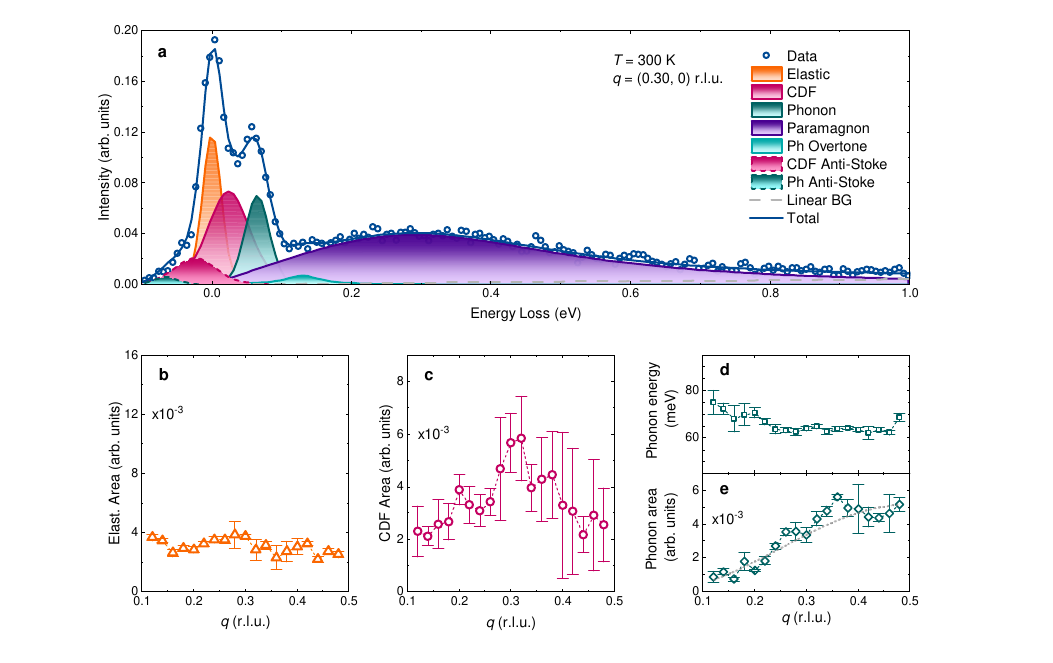}
    \caption{\textbf{Low-energy excitations at 300 K from the fits of the ultra-high-resolution spectra.} \textbf{(a)} Example of fit at $q=(0.3,0)$ r.l.u.; fitting components are the same as in Fig.~\ref{fig:figure2}.  To account for thermal effects, analogous peaks in the anti-Stokes region were also included for the CDF and phonon contributions. From the fits as a function of momentum along the $(H,0)$ direction we extract:\textbf{ (b)} the area of the elastic line, \textbf{(c) }the area of the CDF peak, \textbf{(d)} the dispersion, and \textbf{(e)} the area of the bond-stretching phonons. The CDF intensity is slightly reduced compared to 110 K, and the phonon anomalies are correspondingly less pronounced.}
    \label{fig:fit300}
\end{extfigure}

\begin{extfigure}[H]
    \centering
    \includegraphics[width=1\linewidth]{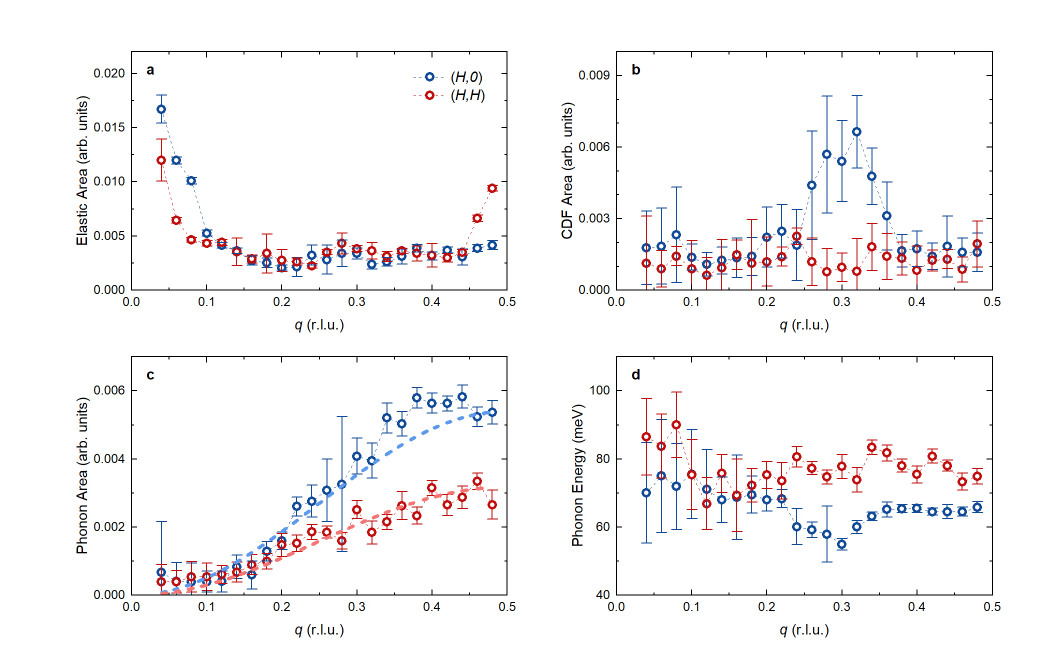}
    \caption{\textbf{Comparison of the low-energy excitations at 110 K between the $(H,0)$ and $(H,H)$ directions.} Ultra-high-resolution spectra along the $(H,H)$ direction were fitted using the same procedure as for the $(H,0)$ direction (Fig.~\ref{fig:figure2}(a)–(d)). A direct comparison between the two directions is shown for (\textbf{a}) the elastic peak, (\textbf{b}) the CDF intensity, (\textbf{c}) the bond-stretching (BS) phonon intensity, and (\textbf{d}) the BS phonon energy. The elastic peak is essentially direction-independent, confirming the negligible presence of static CDW in our sample. By contrast, CDF as well as the anomalies in phonon intensity and energy are clearly visible along $(H,0)$ but vanish along $(H,H)$, highlighting their strong interconnection. In panel (c), the dashed lines indicate the expected $\sin^{2}(\pi q)$ dependence of the BS phonon intensity. 
}
    \label{fig:HH}
\end{extfigure}

\begin{extfigure}[H]
    \centering
    \includegraphics[width=1\linewidth]{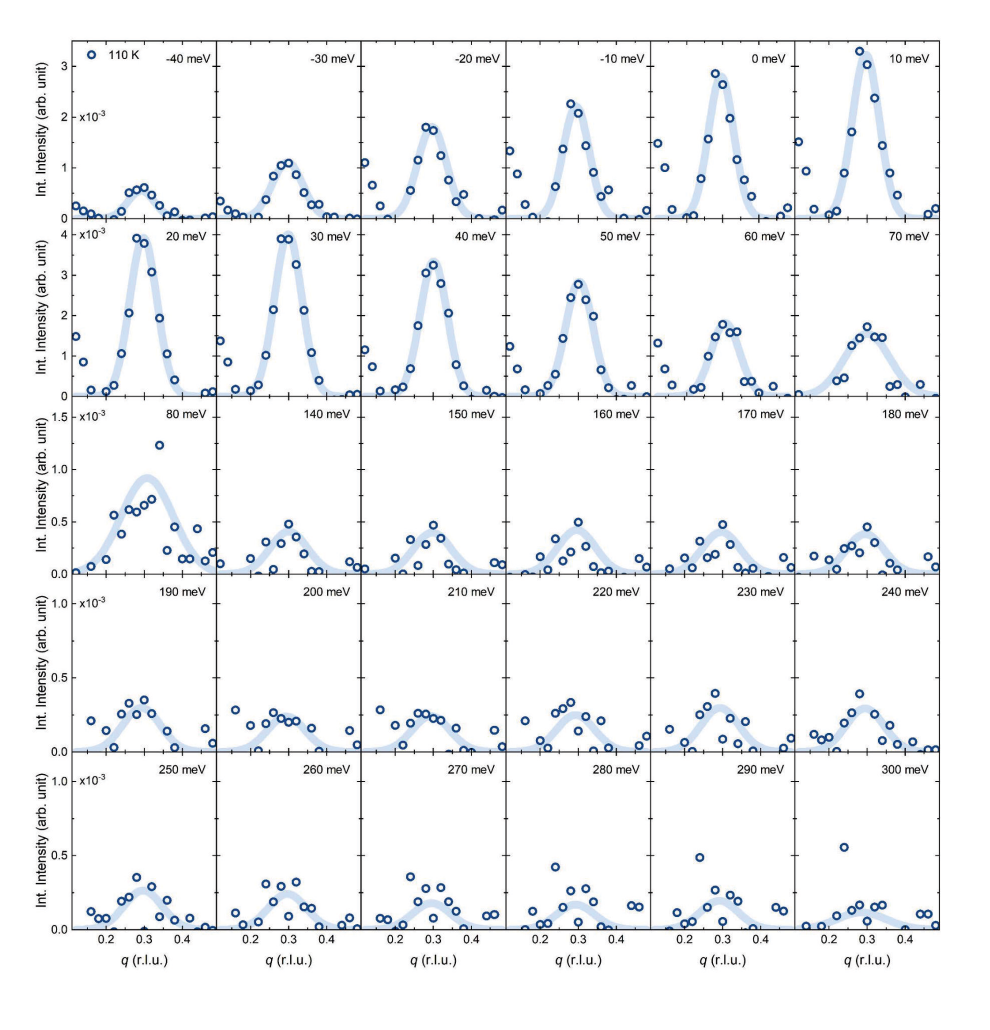}
    \caption{\textbf{Momentum dependence of the integrated ultra-high-resolution spectra at 110 K.} 
    The integrated intensity (circles) is plotted as a function of momentum for each of the 60 meV–wide energy intervals into which the spectra have been divided. The value in the upper-right corner of each panel indicates the center of the corresponding interval. Data are fitted with gaussian profiles (solid lines). For energy intervals centered above 300 meV the integrated intensity becomes indistinguishable from the $q$-dependent background, preventing a reliable gaussian peak determination.}
    \label{fig:Greven110}
\end{extfigure}

\begin{extfigure}[H]
    \centering
    \includegraphics[width=1\linewidth]{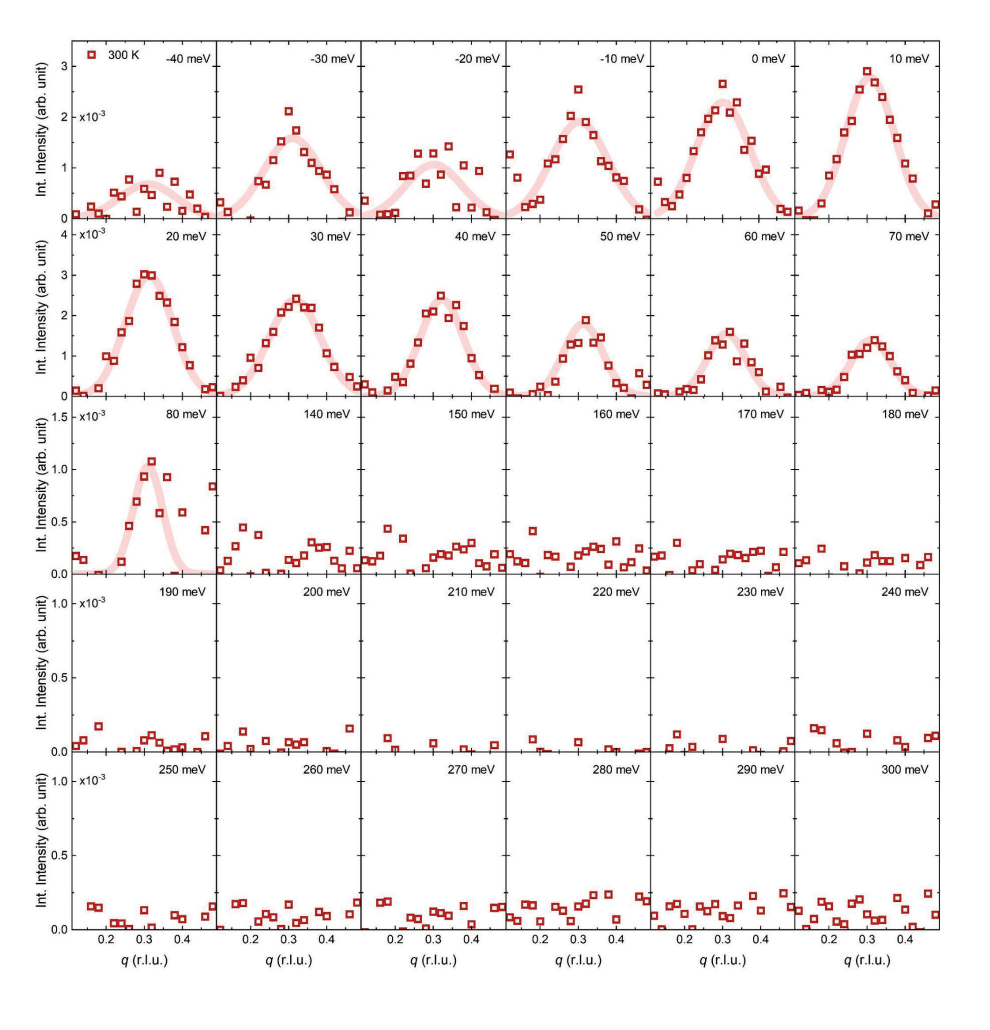}
    \caption{\textbf{Momentum dependence of the integrated ultra-high-resolution spectra at 300 K.} 
    Same analysis as in Supplementary Fig.~\ref{fig:Greven110}, but for spectra measured at 300 K. 
    At this temperature, the quasi-elastic signal rapidly weakens with increasing energy, and already above 80 meV the integrated intensity becomes indistinguishable from the $q$-dependent background, preventing a reliable identification of a CDF peak.}
    \label{fig:Greven300}
\end{extfigure}

\begin{extfigure}[H]
    \centering
    \includegraphics[width=0.8\linewidth]{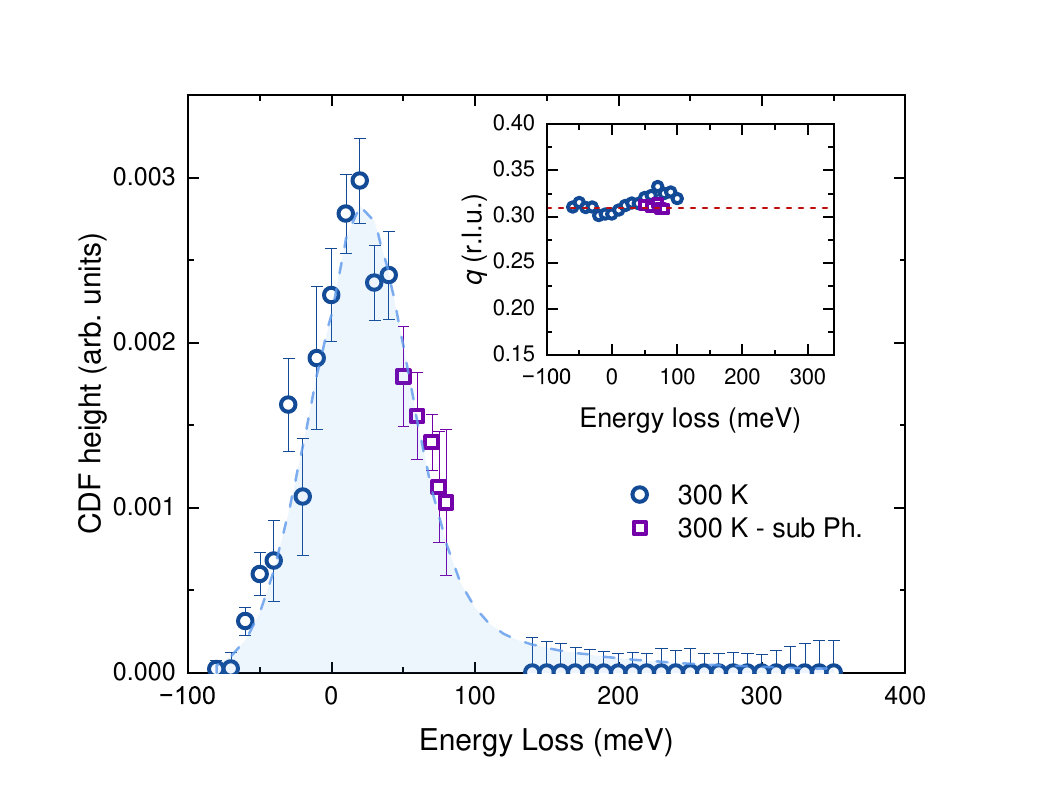}
    \caption{\textbf{Energy profile of CDF at 300 K.} 
    The height of the gaussians determined in Supplementary Fig.~\ref{fig:Greven300} is plotted as a function of energy.  The inset shows the momentum position of the gaussians as a function of energy, which remains aligned with $q_{\mathrm{CDF}}$. Above $\sim$40 meV the signal is dominated by bond-stretching phonons; violet symbols represent the residual CDF component obtained after subtraction of the phonon contribution, both in terms of integrated intensity and momentum position.}
    \label{fig:CDF300}
\end{extfigure}

\begin{extfigure}[H]
    \centering
    \includegraphics[width=1.04\linewidth]{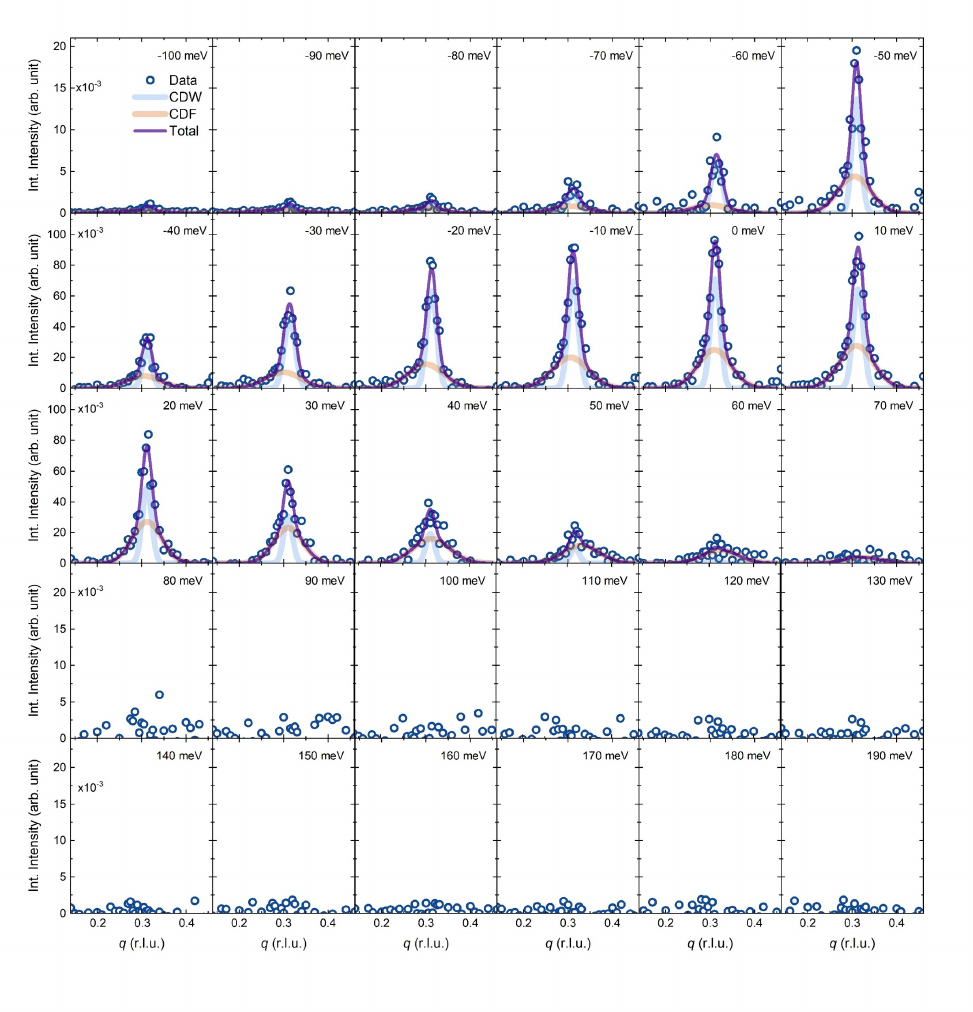}
    \caption{\textbf{Momentum dependence of the integrated spectra of YBCO $p=0.13$.} 
    The integrated intensity at 20 K (circles) is plotted as a function of momentum for each 60 meV–wide energy interval (whose center is reported in the upper right corner of each panel). The signal is described by the superposition of a narrow gaussian (blue line), associated with the quasi-static CDW peak, and a much broader gaussian, characteristic of the short-ranged CDF. 
    Compared to Hg1223, the latter does not display a high-energy extension, highlighting a key difference between the two cuprate families.}
    \label{fig:YBCOUD}
\end{extfigure}

\begin{extfigure}[H]
    \centering
    \includegraphics[width=1.04\linewidth]{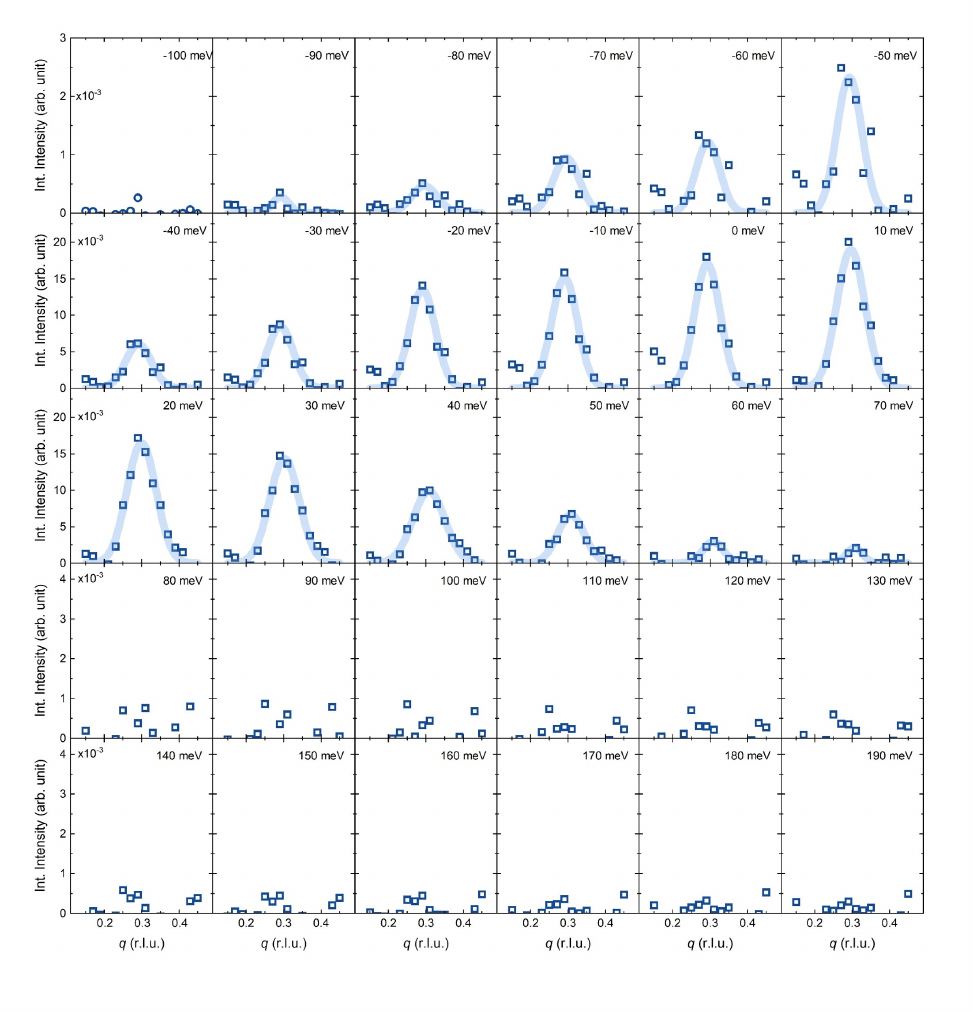}
    \caption{\textbf{Momentum dependence of the integrated spectra of YBCO $p=0.19$.} 
    The integrated intensity at 80 K (squares) is plotted as a function of momentum for each 60 meV–wide energy interval.  As for the underdoped sample (Supplementary Fig.~\ref{fig:YBCOUD}), no significant high-energy extension of the CDF component is observed.}
    \label{fig:YBCOOD}
\end{extfigure}

\begin{extfigure}[H]
    \centering
    \includegraphics[width=1\linewidth]{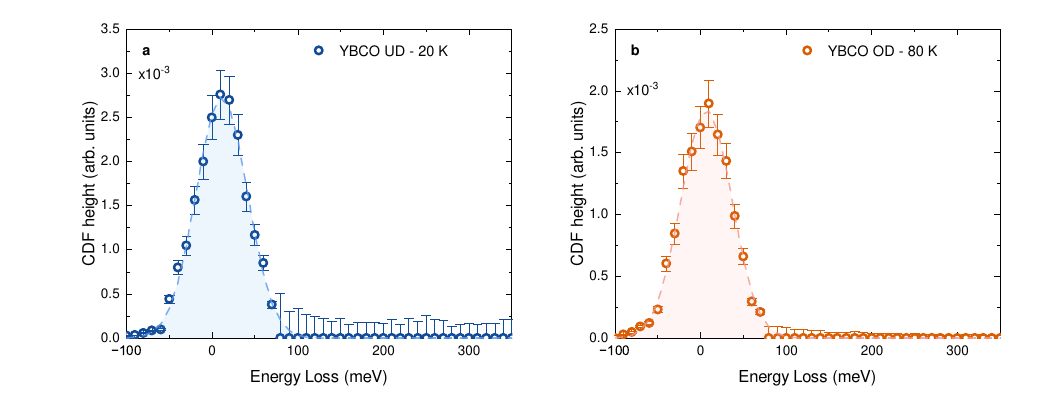}
    \caption{\textbf{Energy profile of CDF in YBCO.} 
    The height of the gaussians extracted \textbf{(a)} from the fits of Supplementary Fig.~\ref{fig:YBCOUD} ($p=0.13$) and \textbf{(b)} Supplementary Fig.~\ref{fig:YBCOOD} ($p=0.19$) is plotted as a function of energy. 
    In both cases the CDF peak is symmetric and confined to the quasi-elastic region.}
    \label{fig:CDFYBCO}
\end{extfigure}

\begin{extfigure}[H]
    \centering
    \includegraphics[width=1\linewidth]{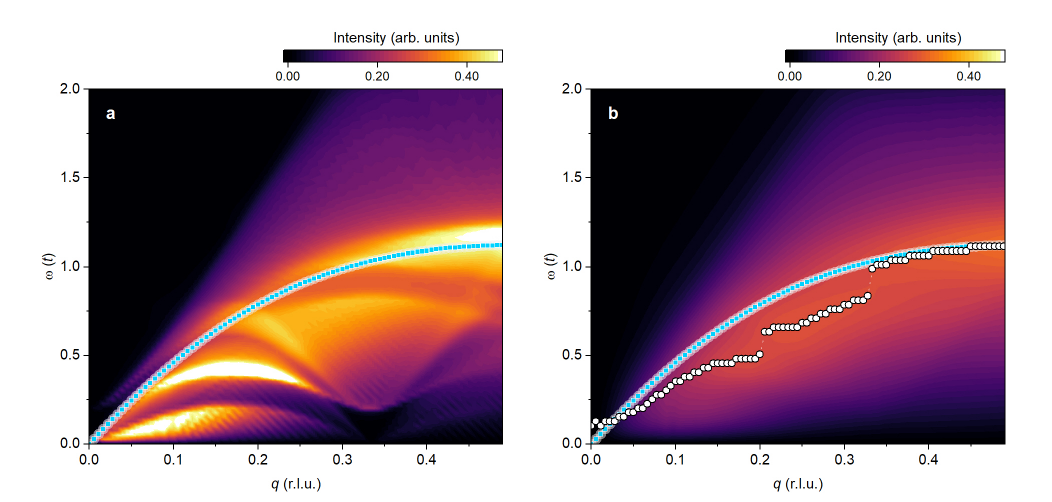}
    \caption{\textbf{Microscopic theory of paramagnon excitations in the presence of CDF.} 
    \textbf{(a)} Paramagnon dispersion for a CDW system with $\Delta/t=0.2$. The energies $\omega$ are measured in units of the nearest-neighbor hopping $t$. \textbf{(b)} Same as in (a), after convolution in frequency with a Lorentzian of width $0.3\,t$, introduced to mimic the broad energy width of the CDF excitations. 
Cyan symbols indicate the paramagnon dispersion of the homogeneous system ($\Delta=0$), while in (b) the white symbols trace the maxima of the convoluted spectral function. 
Other parameters are $t'/t=-0.2$, doping $p=0.15$, $Z=0.5$, and $U/t=4$.}
    \label{fig1M}
\end{extfigure}

\bibliographystyle{unsrtnat} 
\bibliography{Hg1223_ID32}
\end{document}